\documentclass{webofc}

\usepackage[varg]{txfonts}   
\usepackage{hyperref}
\usepackage{url}
\hypersetup{colorlinks=true,citecolor=blue,urlcolor=blue,linkcolor=blue}
\usepackage{lineno}

\begin{document}
\title{Measurement of directed flow of $K^{*0}$ and $\phi$ resonances in Au+Au collisions at RHIC BES energies}
%
%

\author{\firstname{Md } \lastname{Nasim (for the STAR Collaboration)}\inst{1}\fnsep\thanks{\email{nasim@iiserbpr.ac.in}} 
}

\institute{Department of Physical Sciences, Indian Institute of Sciences Education and Research, Berhampur, India-760003 
          }

\abstract{We report measurements of the directed flow ($v_{1}$) for charged kaons, $\phi$ mesons, and $K^{*0}$ resonances in Au+Au collisions at $\sqrt{s_{NN}}$ = 14.5, 19.6, and 27 GeV. This analysis includes the first-ever $v_{1}$ measurement for the $K^{*0}$ resonance in heavy-ion collisions. Our results reveal a centrality-dependent difference in directed flow between charged kaons and $K^{*0}$ resonances, with the difference increasing toward more central collisions. In contrast, the $v_{1}$ difference between kaons and $\phi$ mesons remains nearly constant across centralities. The observed kaon–$K^{*0}$ difference can be qualitatively understood within a hydrodynamic framework that incorporates a hadronic afterburner and an asymmetric loss of $K^{*0}$ yields in momentum-space. Since hadronic rescattering depends strongly on the system size and scattering cross sections among hadrons, the measured $K^{*0}$ $v_{1}$ offers valuable constraints on phase-space–dependent rescattering effects in heavy-ion collisions, thereby providing important input for transport-based models of QCD matter.
}
\maketitle
\section{Introduction}
\label{intro}
High-energy nuclear collisions create extreme conditions of temperature and density, allowing scientists to study how quarks and gluons behave in such environments. These collisions can form a special state of matter, called the quark–gluon plasma (QGP) where quarks and gluons are no longer bound inside protons and neutrons.  As the system expands and cools to near the transition temperature, it changes from QGP into the hadronic phase. At chemical freeze-out (CFO), inelastic collisions stop, locking in the number of each type of particle, while elastic collisions continue until kinetic freeze-out (KFO).
Short-lived resonances are useful tools for studying the hadronic stage in heavy-ion collisions~\cite{star_kstar_bes}. The $K^{*0}$ meson is especially sensitive because its lifetime is similar to the lifetime of the hadronic medium. It decays into a charged kaon and a charged pion, $K^{*0}(\overline{K^{*0}}) \rightarrow K^{\pm}\pi^{\mp}$, with a branching ratio of 2/3. After decay, the daughter particles can scatter with other hadrons, making it harder to reconstruct the original $K^{*0}$, or they can combine to regenerate new $K^{*0}$ mesons through pseudo-elastic interactions. The balance between these two processes affects the $K^{*0}/K$ yield ratio: a decrease with higher multiplicity signals dominant rescattering, while an increase suggests strong regeneration~\cite{star_kstar_bes,kstar_thermal,kstar_jpg}. Experiments show that the $K^{*0}/K$ ratio is smaller in heavy-ion collisions than in smaller systems like $p+p$, consistent with stronger hadronic rescattering in larger systems~\cite{star_kstar_bes}.\\

Collective flow measurements help us understand how the medium created in heavy-ion collisions evolves. The first harmonic term in the Fourier expansion of particle azimuthal distributions, called directed flow ($v_1$), describes the sideward motion of particles and is odd in rapidity. Both hydrodynamic and transport models indicate that $v_1$ is sensitive to the system’s early-stage dynamics as well as late-stage hadronic interactions, making it an important tool to study both the partonic and hadronic phases.
A recent hydrodynamic study with a hadronic phase included via the UrQMD afterburner examined the $v_1$ of $K^{*0}$, $\phi$, and charged kaons~\cite{kstar_hydro}. It found that the $v_{1}$ slope($dv_{1}/dy$) difference between $\phi$ and $K^{+}$ stays consistently negative for all centralities. However, for $K^{*0}$ and $K^{+}$, the $v_1$ difference shows a strong dependence on centrality: it is negative when only primordial $K^{*0}$ mesons are considered, but changes sign once hadronic interactions are included. This sign reversal is linked to asymmetric changes in the $K^{*0}$ yield in both coordinate and momentum space caused by late-stage rescattering. While $K^{*0}$ production yields have been widely measured to study rescattering effects, its directed flow is an even more sensitive probe of the scattering dynamics and dissipative behavior of the hadronic medium.
In this proceeding, we report the first measurement of rapidity-odd directed flow for $K^{*0}$ and $\overline{K^{*0}}$ mesons in Au+Au collisions at $\sqrt{s_{NN}} = 14.5$, 19.6, and 27 GeV, using minimum-bias (MB) data from the STAR experiment collected during the second phase of the Beam Energy Scan (BES-II) program at the Relativistic Heavy Ion Collider (RHIC).

\section{Experiment and methods}
\label{sec-1}
 The STAR detector setup provides uniform acceptance, full azimuthal coverage, and excellent capability for particle identification. For event selection, the primary collision vertex is required to satisfy $|V_z| < 140$ cm (70 cm for 27 GeV) along the beam axis and $|V_r| < 2$ cm in the radial direction. The collision centrality is determined from the multiplicity of raw charged particles measured by the Time Projection Chamber (TPC) detector within the pseudo-rapidity window $|\eta| < 0.5$. Particle identification is carried out using information from both the TPC and the Time-of-Flight (TOF) detectors. The Event Plane Detector (EPD) covering the pseudorapidity range $2.1 < |\eta| < 5.1$ is used to reconstruct the first-order event-plane angle $(\Psi_1)$, while the large $\eta-$ gap between the EPD and the TPC detector helps suppress non-flow contributions in the $v_1$ measurement.
 
 To reconstruct the resonance signals, the combinatorial background is evaluated using a track-rotation technique, where the momentum of one daughter particle is rotated by $180^{\circ}$ in the transverse plane to remove genuine correlations. The raw signal is then obtained by subtracting the combinatorial background from the same-event invariant mass distribution. The directed flow ($v_1$) is extracted using the invariant mass method~\cite{inv_mass}.

\section{Results and Discussion}
\label{sec-2}

\begin{figure}[h]
\centering
\includegraphics[scale =0.43]{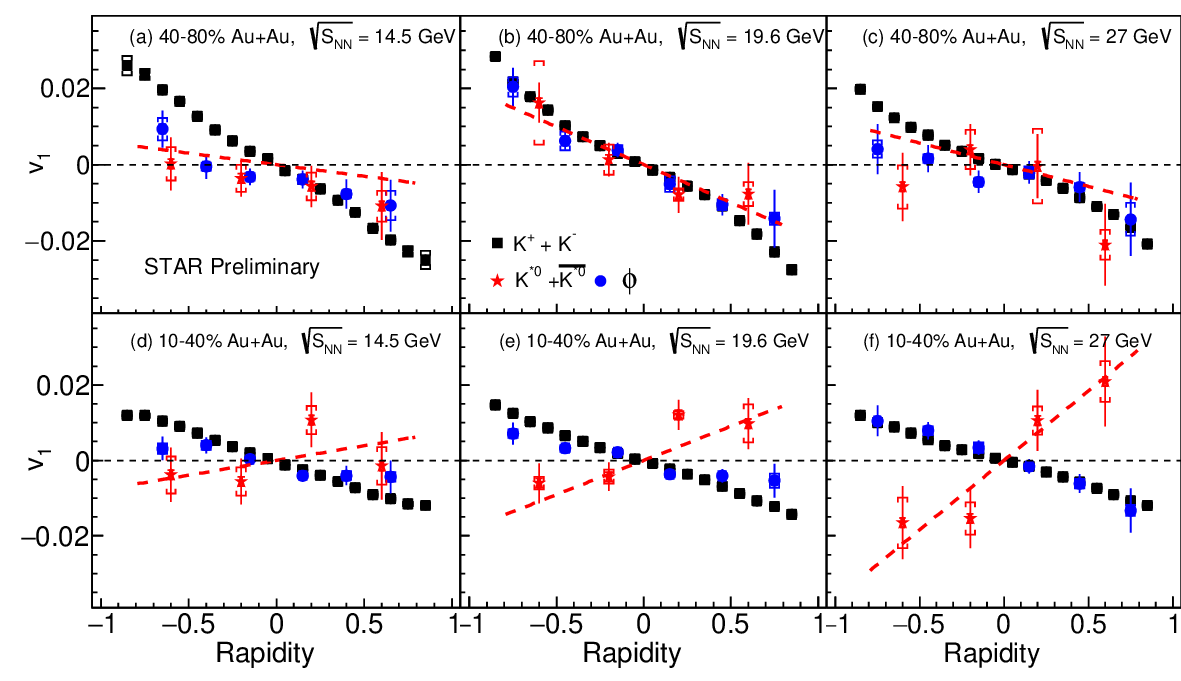}
\caption{ Rapidity (y) dependence of directed flow ($v_1$) for $K^{*0}, \phi~\text{and} ~K$ meson for 10-40\% (lower pannel) and 40-80\% (upper pannel) Au+Au collisions at $\sqrt{s_{NN}}$ = 14.6, 19.6 and 27 GeV.}

\label{fig-1}       
\end{figure}

Figure~\ref{fig-1} presents the $v_1$ as a function of rapidity for $K^{*0}$, $\phi$, and $K$ mesons in 40--80\% (peripheral) and 10--40\% (mid-central) Au+Au collisions at $\sqrt{s_{NN}} = $ 14.6, 19.6, and 27 GeV. The slope of $v_1(y)$ near mid-rapidity is obtained by fitting the distribution with a linear function, $v_1(y) = p_0 y$, restricted to pass through the origin. In peripheral collisions, all three particles exhibit a negative $v_1$ slope. In contrast, for mid-central collisions, the $\phi$ meson and charged kaons continue to show negative slopes, while the $K^{*0}$ meson displays a positive slope. Given that the $K^{*0}$ is a short-lived resonance with a lifetime of about 4.16 fm/$c$, it is more susceptible to in-medium hadronic interactions compared to long-lived resonances such as the $\phi$ meson (lifetime $\approx$ 42 fm/$c$). Hence the observed difference in the $v_1$ slope may therefore be attributed to late-stage hadronic effects. Similar trends have also been predicted by hydrodynamic model calculations~\cite{kstar_hydro}.

\begin{figure*}
\centering
\includegraphics[scale =0.5]{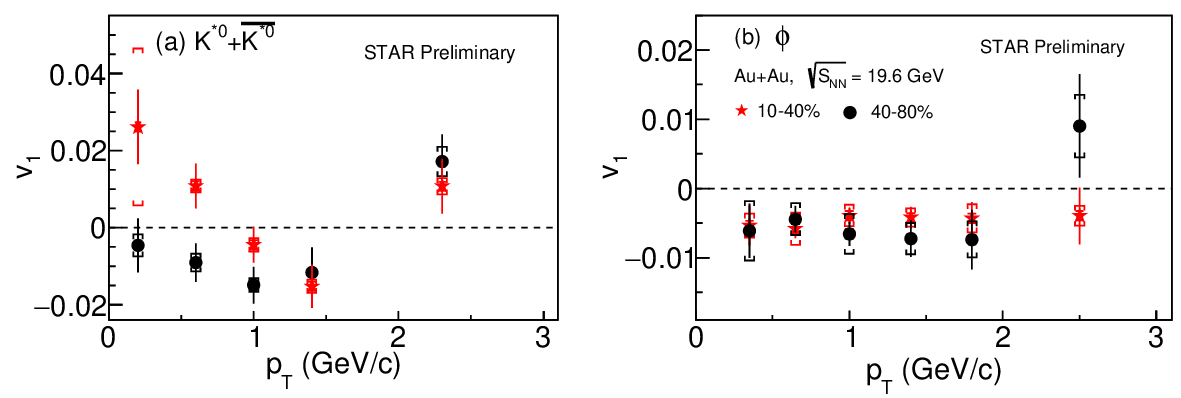}
\caption{ Transverse momentum ($p_{T}$) dependence of directed flow ($v_1$) for $K^{*0}$ (left panel) and $ \phi$ (right panel) meson for 10-40\% and 40-80\% Au+Au collisions at $\sqrt{s_{NN}}$ = 19.6 GeV.}
\label{fig-2}       
\end{figure*}

 Figure~\ref{fig-2} shows the $p_T$-differential, rapidity-integrated $v_1$ for $K^{*0}$ and $\phi$ mesons in peripheral and mid-central Au+Au collisions at $\sqrt{s_{NN}} = 19.6$ GeV. Due to the odd-symmetric definition of $v_{1}^{\text{odd}}$, when presenting the rapidity-integrated $v_1$, the values at negative rapidity are multiplied by $-1$. In mid-central collisions, the $K^{*0}$ meson exhibits a positive $v_1$ at low $p_T$ ($p_T < 1$ GeV/$c$), whereas it remains negative in peripheral collisions. For the $\phi$ meson, $v_1$ stays negative in both centrality classes. At higher $p_T$, both $K^{*0}$ and $\phi$ mesons show positive $v_1$, consistent with earlier $v_1$ measurements for charged hadrons and with expectations from dipolar flow~\cite{bulk_14}. The sign change of $v_1$ in mid-central collisions further supports the interpretation that late-stage hadronic interactions, particularly significant at low $p_T$, play an important role in modifying the $v_1$ slope.

\begin{figure*}
\centering
\includegraphics[scale =0.5]{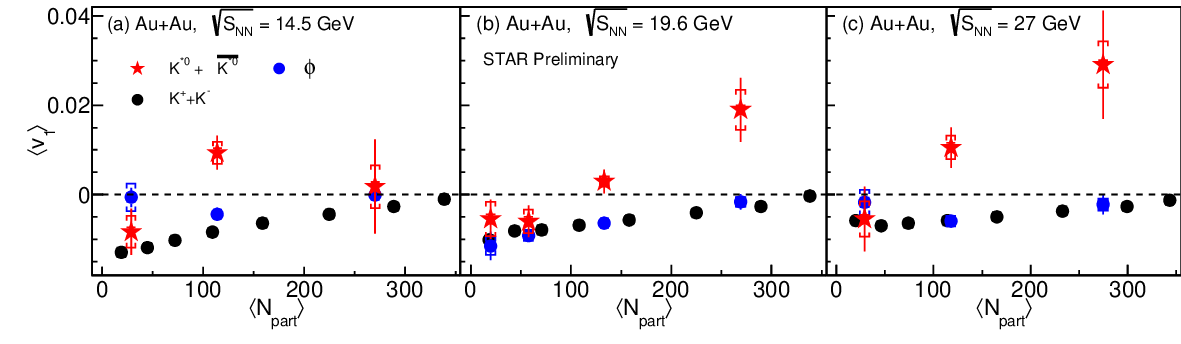}
\caption{Rapidity and $p_{T}$-integrated directed flow ($v_1$) as a function of number of participating nucleons ($\langle N_{part}\rangle$) for $K^{*0}, \phi~\text{and} ~K$ meson in Au+Au collisions at $\sqrt{s_{NN}}$ = 14.6, 19.6 and 27 GeV. }
\label{fig-3}       
\end{figure*}

\begin{figure}
\centering
\includegraphics[scale =0.25]{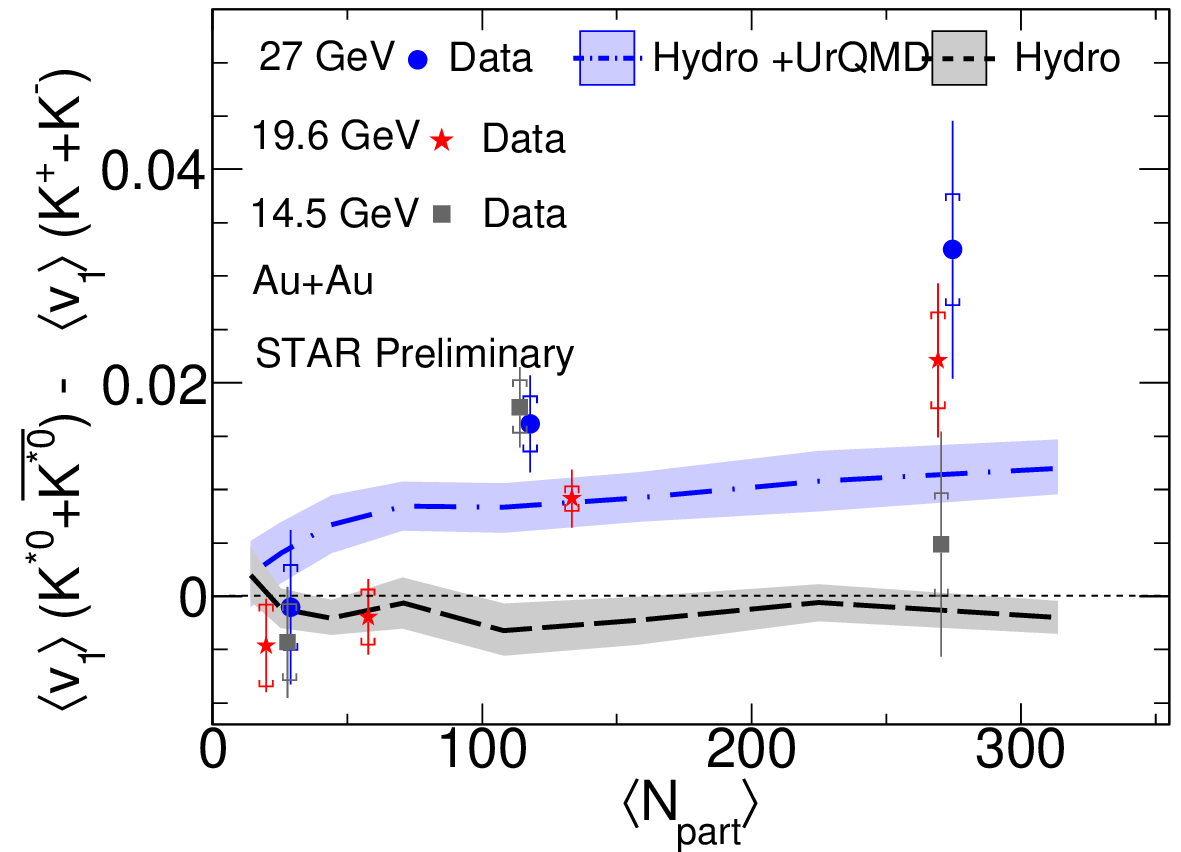}
\caption{The difference between $\langle v_1 \rangle$ of $K^{*0}~\text{and}~K$ meson as a function of number of participating nucleons ($\langle N_{part}\rangle$) in Au+Au collisions at $\sqrt{s_{NN}}$ = 14.6, 19.6 and 27 GeV. The hydrodynamic model calculations are shown with shaded bands.}  
\label{fig-4}       
\end{figure}

Figure~\ref{fig-3} shows the comparison of average $\langle v_1\rangle$ of $K^{*0}$, $\phi$, and $K$ mesons as a function of centrality in Au+Au collisions at $\sqrt{s_{NN}}$ = 14.6, 19.6, and 27 GeV. The $\phi$ meson and charged kaons exhibit a smooth centrality dependence, gradually increasing from peripheral to central collisions while retaining negative values. In contrast, the $K^{*0}$ meson shows a sign change in $\langle v_1\rangle$ when going from peripheral to central collisions. In Fig.~\ref{fig-4}, the difference between the average $\langle v_1 \rangle$ of $K^{*0}$ and $K$ mesons is shown as a function of $\langle N_{part}\rangle$, together with comparisons to hydrodynamic model calculations~\cite{kstar_hydro}. In the model results, the difference in $\langle v_1\rangle$ is found to be close to zero when the hadronic phase is not included. However, after incorporating the UrQMD afterburner, the model predictions reasonably describe the data. This suggests that the centrality-dependent difference in $\langle v_1\rangle$ between $K^{*0}$ and $K$ mesons might originate from the  asymmetric loss of $K^{*0}$ mesons due to rescattering during the hadronic phase evolution.

\subsection{Summary}
The first measurement of directed flow for $K^{*0}$ mesons is presented in Au+Au collisions at $\sqrt{s_{NN}} =$ 14.6, 19.6, and 27 GeV. The results are compared with that of $\phi$ mesons and charged kaons. The $v_{1}$ of $K^{*0}$ exhibits a sign change from peripheral to mid-central collisions. The centrality-dependent difference in the average $\langle v_1 \rangle$ between $K^{*0}$ and $K$ mesons increases from peripheral to central collisions, a trend that is well reproduced by hydrodynamic model calculations including the UrQMD afterburner. This measurement sheds light on the phase-space–dependent asymmetric loss of short-lived resonances during the hadronic phase evolution in heavy-ion collisions.

%
%
%

\end{document}